\documentclass[prl,floatfix,showpacs,amsmat,twocolumn]{revtex4}
\usepackage{amssymb}
\usepackage{amsmath}
\usepackage{graphicx}
\usepackage[colorlinks]{hyperref} 

\newcommand{\ket}[1]{|#1\rangle}
\newcommand{\bra}[1]{\langle#1|}

\newcommand{\bl}{\begin{itemize}}
\newcommand{\el}{\end{itemize}}

\newlength{\myL}

\begin{document}
\title{Measurement-based quantum computer in the gapped ground state \\ 
of a two-body Hamiltonian}
\author{Gavin K. Brennen$^{1,2}$ and Akimasa Miyake$^{2,3}$}
\affiliation{%
\mbox{$^1$ Department of Physics, Macquarie University, Sydney, NSW 2109, 
Australia} \\
\mbox{$^2$ Institute for Quantum Optics and Quantum Information,
Austrian Academy of Sciences, Innsbruck, Austria} \\ 
\mbox{$^3$ Institute for Theoretical Physics, University of Innsbruck,
Technikerstrasse 25, A-6020 Innsbruck, Austria}
}
\date{March 10, 2008} 

\begin{abstract}
We propose a scheme for a ground-code measurement-based quantum computer,
which enjoys two major advantages.  First, every logical qubit is encoded
in the {\it gapped} degenerate ground subspace of a spin-$1$ chain with 
nearest-neighbor {\it two-body} interactions, so that it equips built-in
robustness against noise.
Second, computation is processed by single-spin measurements along multiple 
chains dynamically coupled on demand, so as to keep teleporting only logical 
information into a gap-protected ground state of the residual chains after 
the interactions with spins to be measured are turned off.
We describe implementations using trapped atoms or polar molecules in an 
optical lattice, where the gap is expected to be as large as $0.2$ kHz 
or $4.8$ kHz respectively.
\end{abstract}

\maketitle

{\bf Introduction.---}  
Reliable quantum computers require hardware 
with low error rates and sufficient resources to perform software-based error 
correction.  One appealing approach to reduce the massive overhead for error 
correction is to process quantum information in the {\it gapped} ground states 
of some many-body interaction.  This is the tactic used in topological quantum 
computation and adiabatic quantum computation.
Yet, the hardware demands for the former are substantial, and the 
fault tolerance of the later, especially when restricted to two-body 
interactions, is unclear \cite{Jordan:06}.
On the other hand, measurement-based quantum computation (MQC), in particular
one-way computation on the 2D cluster state \cite{Raussendorf:01}, runs 
by beginning with a highly entangled state dynamically generated 
from nearest-neighbor two-body interactions and performing computation by only 
single-qubit measurements and feed forward of their outcomes.   
However, its bare implementation may suffer decoherence of physical qubits 
waiting for their round of measurements in the far future, and that severely 
damages a prominent capability to parallelize computation. 
Although its fault-tolerant method by error correction has 
been well established \cite{FTMQC}, it is clearly advantageous if some 
gap-protection is provided on the hardware level.

In this paper, we propose a ground-code measurement-based
quantum computer (GMQC), which enjoys the two aforementioned advantages.  
GMQC is a conceptual advance, since measurements generally create 
excitations in the system so that two desired properties, keeping 
the information in the ground state and processing the information by 
measurements, are not seemingly compatible.
We demand three properties of the ground state for GMQC.
(i) There should be an energy gap to penalize errors 
moving outside the computational ground subspace, and operators
connecting logical states should be highly nonlocal.  These should persist 
in the thermodynamic limit to be scalable.
(ii) The interactions should be preferably two-body.
It is possible that in some ground subspace of $H$ the effective interaction 
is $K$-local, but a demerit is the significantly reduced magnitude of 
the perturbative coupling.
(iii) The interactions should be frustration-free,
so that when every single spin is measured through computation, the remaining 
entangled spins to contain logical information can be set in the ground 
state of their Hamiltonian.

We briefly refer to entangled resource states universal for MQC in the 
literature.
First, the idea to use a ground state of the two-body Hamiltonian for 
universal MQC is seen in Ref.~\cite{Bartlett:06}, where 
the effective five-body Hamiltonian for the 2D cluster state is perturbatively 
approximated in a low energy sector using ancillas.
We do not resort to perturbation, so that not only is the gap much 
larger in practice, but also the resource state is exact in that 
we could approach unit fidelity as close as possible, by improving accuracy of 
analog simulation of our Hamiltonian and its preparation to the ground state.
Second, the novel use of tensor network states beyond the 2D cluster 
state \cite{Gross:07a,Gross:07b} has been quite motivating for us. But, our key
idea of processing the logical information in degenerate ground states while
maintaining the Hamiltonian on is incorporated for the first time 
in our paper.

\begin{figure}[t]
\begin{center}
\includegraphics[width=1.00\columnwidth]{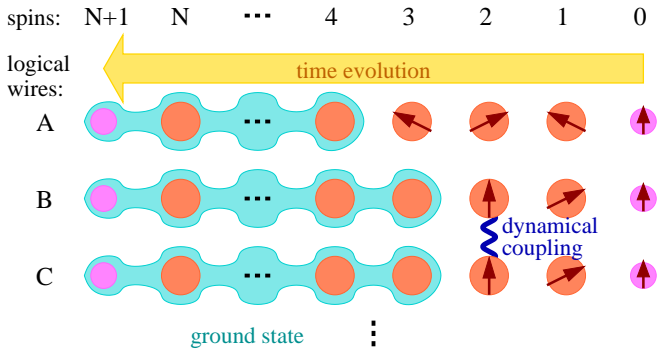}
\end{center}
\caption{A schematic of a ground-code measurement-based quantum computer.  
Each logical qubit is encoded in every AKLT chain of $N$ spin-$1$ particles 
with spin-$1/2$ at the boundaries $0$ and $N+1$, where spin particles
interact with the nearest neighbors.
The initial state of the chain is a unique ground state with 
a constant gap, and the measurement of the $0$-th spin prepares the logical wire
in $\ket{0^L}$ or $\ket{1^L}$. 
Thereafter, computation is carried by measurements right to left, in processing
the logical information in the two-fold degenerate, gapped ground state of
each residual chain.
Every single-qubit gate on the wire $A$ is performed non-deterministically 
at the time step $j$ by turning off the interaction $P^2_{A_j,A_{j+1}}$ and 
measuring the spin $A_j$ in an appropriate basis.  A non-deterministic  
two-qubit gate, ${\tt CPHASE}_{B,C}$, at the time step $j$ is done by turning off 
both $P^2_{B_j,B_{j+1}}$ and $P^2_{C_j,C_{j+1}}$, and then applying a dynamical 
coupling  $\exp (i\pi H^{\rm int}/\chi)$ followed by measurements of the 
spins $B_j,C_j$.
Gate failures, which occur with bounded probabilities, are heralded, and 
act just as the logical identity by the adaption of subsequent measurement bases.
The final readout is made by measuring the spin-$1/2$ at $N+1$.}
\label{fig:scheme}
\end{figure}

{\bf Scheme of a ground-code MQC.---}
In our GMQC, sketched in Fig.~\ref{fig:scheme}, we adopt 
a hybrid approach where the logical two-qubit gates are implemented via 
dynamical couplings on demand as in the quantum circuit model, 
in addition to the standard MQC for the time evolution of each logical qubit.  
This is partly because there has not been known any exact gapped ground state 
of a two-body Hamiltonian which per se serves as a universal resource for 
the standard MQC.
We utilize space-time resources in such a joint way that a
 ``spatially'' entangled resource is used to simulate the logical time 
evolution and ``temporal'' interaction is used to simulate the logical 
spatial interactions.
Consequently, our GMQC exhibits some new features in contrast with the 
conventional MQC.

%
%
The basic Hamiltonian we consider is the 1D Affleck-Kennedy-Lieb-Tasaki
(AKLT) model \cite{AKLT:88}, the chain of nearest-neighbor two-body 
interacting spins,
$H=J[\sum_{j=1}^{N-1}P^2_{j,j+1} +P^{3/2}_{0,1}+P^{3/2}_{N,N+1}] $,
with $J>0$ and $P^S_{j,j+1}$ the projector onto the spin-$S$ irreducible 
representation of the total spin for particles $j$ and $j+1$.
Namely, $P^2_{j,j+1} = \frac{1}{2}({\bf S}_{j} \cdot {\bf S}_{j+1} + 
\tfrac{1}{3} ({\bf S}_{j} \cdot {\bf S}_{j+1})^2) + \frac{{\bf 1}_9}{3}$ and 
$P^{3/2}_{j,j'} = \tfrac{2}{3}({\bf 1}_6 + {\bf s}_{j} \cdot {\bf S}_{j'}),$ 
where ${\bf S},{\bf s}$ are spin-$1$, $1/2$ representations of 
$\mathfrak{su}(2)$.
Without the boundary spins, the finite AKLT spin chain has a four-fold 
degeneracy corresponding to a total spin-$0$ (singlet) state and 
a triplet of spin-$1$ states. The boundary terms $P^{3/2}$
project out the spin-$1$ components yielding a unique ground state
such that $H\ket{\mathcal G}=0$. 
The gap $\Delta E$ persists in the thermodynamic limit,
and has been estimated to be $\Delta E\approx 0.350 J$ \cite{Werner:91}.  
A key feature of $\ket{\mathcal G}$ is that it can serve, using
{\it single-spin} measurements only, as a logical quantum wire which is capable 
of performing not only deterministic teleportation as already remarked in
Ref.~\cite{Verstraete:04a} but also an arbitrary logical single-qubit 
operation similarly to one-way computation as shown below.

The ground state $\ket{\mathcal G}$ has a convenient matrix product state (MPS)
representation \cite{Verstraete:04a,Fan:04}.
Let us define 
$\ket{1_j} = \tfrac{-1}{\sqrt{2}}(\ket{S^z_j = 1} - \ket{S^z_j = -1})$,
$\ket{2_j} = \tfrac{1}{\sqrt{2}}(\ket{S^z_j = 1} + \ket{S^z_j = -1})$, and
$\ket{3_j} = \ket{S^z_j = 0} $, in terms of the three eigenstates of $S^z_j$,
\begin{equation}
\ket{\mathcal G} = \sum_{\{\alpha_j\}=1}^{3} 
\frac{\ket{\alpha_1} \ldots \ket{\alpha_N}}{\sqrt{3^{N}}}
\left[{\bf 1}_2\otimes \prod_{j=N}^1 \bra{\alpha_j} M_j\rangle\right]
\ket{\Psi^{-}_{0,N+1}} ,
\label{eq:ground_AKLT}
\end{equation}
where $\ket{M_j}=X\ket{1_j}-iY\ket{2_j}+Z\ket{3_j}$ ($X,Y,Z$ are the Pauli
matrices $\sigma^{\mu} \; (\mu=x,y,z)$), and 
$\ket{\Psi^{-}}$ is the singlet ($S=0$) located on the $0$-th 
and $N+1$-th sites.
This representation is helpful to see the action of local measurements
to the ``relative'' state of unmeasured parts in simulating unitary 
evolution of MQC \cite{Verstraete:04b,Gross:07a,Gross:07b}.   
Note that, in Refs.~\cite{Gross:07a,Gross:07b}, 
the ground state of a modified AKLT chain, different from 
the original, was considered to construct a resource state.
However, their extension of the so-called byproduct operators into a non-Pauli
finite group is less convenient, and it is unclear whether it still 
possesses a counterpart of the nonlocal string order utilized later.

{\bf Universal computation by measurements.---}
Computation follows by measuring right to left 
(as in reading the Japanese comics)
the single spins along each chain, where spins are indexed by 
increasing value moving right to left, in accordance with 
the order of unitaries we simulate in the bra-ket notation.  
First we need to prepare the unique ground state $\ket{\mathcal G}^{\otimes n}$
of the $n$ parallel decoupled 1D AKLT chains.   
This can be done efficiently and deterministically along each chain, 
by either turning on $H$ immersing the system in 
a reservoir and cooling it, or by making use of its MPS description to 
produce it via sequential unitaries (which scale linearly in $N$) 
\cite{Schon:07} before turning on $H$.

The initialization of every quantum logical wire
is done by first turning off the coupling $P^{3/2}_{0,1}$ and 
measuring the rightmost $0$-th spin-$1/2$ in the $\hat{z}$ basis.
Because of the singlet configuration 
$\ket{\Psi^{-}_{0,N+1}}$, the $-1/2$ and $1/2$ outcome, denoted by $\ket{1_0}$
and $\ket{0_0}$ respectively, at the $0$-th site 
$\hat{z}$-basis measurement induces the initialization of the wire 
(identified effectively with the preparation of the state at the $N+1$-th site,
since we will see that unitary actions accumulate on this degree of freedom) 
to $\ket{0^L}$ and $\ket{1^L}$, respectively.
With one boundary spin-$1/2$, the ground state is two-fold degenerate
spanned by $\langle 1_0 \ket{\mathcal G}$ and 
$\langle 0_0 \ket{\mathcal G}$,
and computation takes place in this ground subspace after this initialization.

%
%
Every logical single-qubit unitary operation is implemented by the single-spin
measurement.
The interaction $P^{2}_{j,j+1}$ is turned off before the local 
measurement of the $j$-th spin, to guarantee that 
the remaining system stays in the ground state.
Since an arbitrary single-qubit operation is decomposed into three rotations 
around the logical $Z$ and $X$ axes with three Euler angles,
we show their measurement directions.
The rotation $R^{Z}(\theta) = \ket{0^L}\bra{0^L} + 
e^{i\theta}\ket{1^L}\bra{1^L}$ along the $Z$ axis is applied by 
the single-spin measurement in an orthogonal basis,
\begin{equation}
\{\ket{\gamma^{Z}_j(\theta)}\} =
\{\tfrac{1}{2}((1\pm e^{-i\theta})\ket{1_j} + (1\mp e^{-i\theta})\ket{2_j}), 
\ket{3_j}\} .
\end{equation}
If the outcome is either the first or the second (which turns out to 
occur always with probability $1/3$ for each), we can apply 
$R^{Z}(\theta)$ newly on the logical qubit with a byproduct operator
$X$ or $XZ$, respectively. If the outcome is the third, we interpret we 
have applied the ``logical identity'' with a byproduct $Z$.
On the other hand, the rotation $R^{X}(\theta) = \ket{+^L}\bra{+^L} + 
e^{i\theta}\ket{-^L}\bra{-^L}$ along the $X$ axis, where 
$\ket{\pm^L} = \tfrac{1}{\sqrt{2}}(\ket{0^L}\pm \ket{1^L})$,
is applied by the single-spin measurement in another orthogonal basis,
\begin{equation}
\{\ket{\gamma^{X}_j (\theta)}\} =
\{\tfrac{1}{2}((1\pm e^{i\theta})\ket{2_j} + (1\mp e^{i\theta})\ket{3_j}), 
\ket{1_j}\} .
\end{equation}
If the outcome is either the first or the second, we apply 
$R^{X}(\theta)$ with a byproduct $XZ$ or $Z$, respectively, and otherwise 
the logical identity with a byproduct $X$.

Suppose we initialize the logical wire $\ket{0^L}$ by
the $0$-th site $-1/2$ outcome (otherwise, we consider the wire is 
initialized $\ket{0^L}$ with the byproduct $X$ from the beginning). 
According to Eq.~(\ref{eq:ground_AKLT}), before the $j$-th spin is measured, 
we have a state 
$\ket{\psi(j)} = \bra{1_0}\prod_{k'=j-1}^{1}
\langle\gamma_{k'}\ket{\mathcal G}$
given by
\begin{equation}
\sum_{\{\alpha_{k} \}=1}^{3}
\frac{\ket{\alpha_{j}} \ldots \ket{\alpha_{N}}}{\sqrt{3^{N-j+1}}}
 \left[\prod_{k=N}^{j}  \bra{\alpha_k} M_k\rangle  
\Upsilon  \prod_{k'=j-1}^{1} R(\theta_{k'})\right]\ket{0^L} ,
\label{eq:interm}
\end{equation}
where the measurement directions of $\ket{\gamma_{k'}(\theta_{k'})}$ must be
adapted from $\theta_{k'}$ to $-\theta_{k'}$ when non-commuting byproducts from 
previous measurements are propagated left through the current one,
resulting in an accumulated byproduct operator $\Upsilon$.

We describe important properties of the residual Hamiltonian
$H(j)=J [\sum_{k=j}^{N-1} P^2_{k,k+1} +P^{3/2}_{N,N+1}]$ 
through the measurement stage of the $j$-th spin.  First, $H(j)$ is gapped 
as before and is two-fold degenerate.  Defining the string operators
$\Sigma^{\mu}(j)
=e^{i\pi \sum_{k=j}^N S^{\mu}_k}\otimes \sigma^{\mu}_{N+1}$,
we find that $[\Sigma^{\mu}(j),H(j)]=0$ whereas 
$\{\Sigma^x(j),\Sigma^z(j)\}_{+}=0$.
At each stage $j$, the pair $\Sigma^{x,z}(j)$ forms
a representation of $\mathfrak{su}(2)$, and
the degenerate ground states are only connected by nonlocal operators.
For every single quantum wire, we utilize ``time-dependent'' logical encoding 
such that $\bra{\psi(j)} \Sigma^{\mu} (j) \ket{\psi(j)} = 
\bra{0_{N+1}}u^{\dagger} \Upsilon^{\dagger} \sigma^{\mu}_{N+1} 
\Upsilon u\ket{0_{N+1}}$, where $u$ is the total single-qubit rotation 
until the $j-1$-th gate.

Second, the logical state is not disturbed when turning off
the interaction coupling the bulk to the $j$-th spin.  
We can decouple with the time-dependent Hamiltonian
$H(j;t)=J (1-c(t))P^{2}_{j,j+1}+H(j+1)$ ,
where $c(t)$ is monotonically increasing in $t \in [0,1]$ with 
$c(0)=0$, $c(1)=1$. Now $P^2_{j,j+1}$ does not commute with $H(j+1)$.  
However, the AKLT Hamiltonian has the property that the ground states 
also minimize its positive summands, i.e., it is frustration-free 
\cite{AKLT:88}. Hence,
$P^{\rm gr}(0)= P^{\rm gr}(t)$ $\forall t < 1$ and 
$\frac{\partial^{\nu} H(j;t)}{\partial t^{\nu}}P^{\rm gr}(t)=0$ 
$\forall \nu, t$, 
where $P^{\rm gr}(t)= \ket{{\mathcal G}(t)}\bra{{\mathcal G}(t)}$ are 
projectors onto the ground subspaces of $H(j;t)$.  Thus, turning off the end 
interaction term does not couple to excited states, and can be done in
a constant time independent of the system size.
Even if there are some unwanted initial perturbations on $H(j;t)$, the gap 
provides robustness if performed adiabatically.
The same argument applies to turning off the boundary terms.

Third, notice that $\ket{\psi (j+1)}$ of Eq.~(\ref{eq:interm}) can be written 
for the general outcome $\ket{r_0}$ ($r = 0,1$) of the $0$-th site measurement 
as
$\tfrac{1}{\sqrt{3^{N-j}}} \sum_{\{\alpha_{k} \}=1}^{3}
\ket{\alpha_{j+1}}\ldots \ket{\alpha_{N}} 
\bra{r_0} [V^{\dagger} \otimes \prod_{k=N}^{j+1}\bra{\alpha_k} M_k\rangle]  
\ket{\Psi^{-}_{0,N+1}}$,
where $V$ is the form of $\tilde{\Upsilon} \prod_{k'=j}^{1}
R(\tilde{\theta}_{k'})$ due to the invariance of $\ket{\Psi^{-}}$ under 
the bilateral unitaries.
But this is equivalent to the state obtained by beginning in the unique ground 
state of $H(j+1)+P^{3/2}_{0,j+1}$, turning off 
$P^{3/2}_{0,j+1}$, measuring the $0$-th spin in the basis 
$\{V\ket{r_0},VX\ket{r_0}\}$, with the result $V\ket{r_0}$.  
This state is in the ground subspace of $H(j+1)$.

%
%
Logical two-qubit operations are made {\it dynamically} by coupling two 
spin-1 particles in adjacent chains, say $A$ and $B$, followed by their 
local measurements (equivalently, by a two-spin measurement). 
First $P^2_{A_{j},A_{j+1}}$ and $P^2_{B_{j},B_{j+1}}$ are turned off. 
We introduce the physical interaction $\exp (iH^{\rm int} \pi/\chi)$
between spins $A_j$ and $B_j$, where 
$H^{\rm int} =  \chi \ket{S^{z}_{A_j}=1}\bra{S^{z}_{A_j}=1}\otimes 
 \ket{S^{z}_{B_j}=1}\bra{S^{z}_{B_j}=1} $,
and measure both spins $A_j$ and $B_j$ in the standard basis 
$\{ \ket{1},\ket{2},\ket{3} \}$.
If the both outcomes are in either $\ket{1}$ or $\ket{2}$, which occurs
with overall probability $(2/3)^2 = 4/9$, we successfully apply the logical
Controlled-Phase gate ${\tt CPHASE}_{A,B} = {\bf 1}_4 - 
2 \ket{1^L_A 1^L_B}\bra{1^L_A 1^L_B}$.
Notice that, in the span by $\ket{1_{A_j} 1_{B_j}}$,
$\ket{1_{A_j} 2_{B_j}}$, $\ket{2_{A_j}1_{B_j}}$, and $\ket{2_{A_j}2_{B_j}}$,
$\exp(iH^{\rm int} \pi/\chi)$ acts as 
$\Gamma = {\bf 1}_4- \tfrac{1}{2}({\bf 1}_2-X)\otimes ({\bf 1}_2-X)$
and as the identity elsewhere.
We see that this induces $\bra{\alpha_{A_j}}\otimes \bra{\beta_{B_j}}\Gamma 
\ket{M_{A_j}}\otimes\ket{M_{B_j}} = \Upsilon {\tt CPHASE}_{A, B}$
with the byproduct $\Upsilon = XZ \otimes XZ$, 
$XZ \otimes X$, $X \otimes XZ$, or $X \otimes X$ in the aforementioned span.
Otherwise (i.e., if at least one outcome is in $\ket{3}$), we end up with
applying the logical identity with the byproduct 
$\bra{\alpha_{A_j}}M_{A_j}\rangle \otimes \bra{\beta_{B_j}}M_{B_j}\rangle$.

To prove that computation is kept in the ground subspace,
imagine that we began in a separable state of two chains $A$ and $B$ 
initialized in $\ket{0^L}$.
After {\tt CPHASE} is successfully applied at the stage $j$ and byproducts 
for $A$ and $B$ are propagated, the joint state involves newly
the operators $(\Upsilon_{A} \otimes \Upsilon_{B})
\tfrac{1}{2}({\bf 1}_4 +Z_{A_{N+1}}+Z_{B_{N+1}}-Z_{A_{N+1}}Z_{B_{N+1}})$,
so that it is nothing but a superposition of logical states each of which 
is in the kernel of $[H_A(j+1)+H_B(j+1)]$ and hence is in the ground subspace 
of the two chains.

It can be verified that for any quantum circuit realized with our universal 
set of gates, the probability for each successful single-, two-qubit 
gate is constant at $2/3$, $4/9$, respectively.  
We can efficiently perform the entire computation, by trying every gate until 
success, at the same time deterministically teleporting (by the standard
basis measurement) other logical qubits to be spatially aligned for 
subsequent two-qubit gates.  
A remarkable new feature is that the general single-, two-qubit operations are 
probabilistic, while the logical identity (teleportation) is essentially 
deterministic with the adaptive measurements. 
This variable computational depth originates from the ``correlated logical
times'' due to a two-point spatial correlation of the AKLT chain which decays 
as $(-1/3)^{|j-j'|}$ between sites $j$ and $j'$.

%
%
At the end of computation, the joint state of leftmost boundary
spins is $\Upsilon U\ket{0^L}^{\otimes n}$, where 
$\Upsilon$ is the total byproduct and $U$ the target unitary operator.
The final logical measurement, without loss of generality in the computational
basis, is simulated by measuring in $\hat{z}$ these spins (after
teleporting the logical information if the chain is redundant).
There, for example, the logical $\ket{0^L}$ outcome must be considered to 
occur if either $s^z_{N+1} = 1/2$ and $\Upsilon$ 
does not contain $X$, or $s^z_{N+1} = -1/2$ and $\Upsilon$ contains $X$.

{\bf Physical implementations.---}
While the scheme of GMQC works independent of implementations, 
we sketch two physical realizations.
In Refs.\cite{Yip:03, Cirac:04}, it was shown how to obtain the AKLT model 
(without the boundary terms) using tunneling induced collisions between 
bosonic atoms trapped one per site in a 3D optical lattice. 
The lattice beams are chosen such that the confinement is very strong along
 $\hat{x},\hat{y}$ directions but weaker along $\hat{z}$, so that the system 
describes non-interacting 1D chains.  
The spin states correspond to $F=1$ hyperfine ground states of an alkali atom, 
and the effective coupling scales like $J\sim t^2/U_S$ where $t$ is the 
tunneling matrix element and $U_S\ (S=0,2)$ is a total spin dependent 
repulsive s-wave scattering energy in a common well.  
For realistic lattice parameters $U_S\approx 5$ kHz and for $t/U_S=0.3$, 
we expect the gap $\Delta E\approx 0.2$ kHz.  
A measurement on spin $A_j$ is achievable by adiabatically expanding 
the lattice along $\hat{z}$ using the accordion technique \cite{Porto:03} 
which decreases $t,U_S$ but also allows the lattice wells within each 
wire to be better addressed.  
Then turning on optical tweezers \cite{Zhang:06} near the site $A_j$, 
one can shift the potential depth at $A_j$ so that tunneling to the 
neighboring wells is effectively zero. State dependent measurements can be 
done using spatially resolving microwave or Raman fields. 
To perform the {\tt CPHASE} gate between $A_j,B_j$ with wires separated along 
$\hat{y}$, after turning off  the interactions apply Raman pulses resonant 
at those locations that map $\ket{S^z=1}$ to the first vibrational 
state of each well.  Introducing a second lattice along $\hat{y}$ with 
twice the period of the first as in Ref.~\cite{Cirac:04}, the intensity can be 
adjusted so that the first excited vibrational states of nearest neighbor wells
interact to generate a tunneling phase gate \cite{Strauch:07}. 
The second lattice intensity can be then tuned to turn off tunneling and 
the states mapped back to vibrational ground states to commence with 
measurement.  To engineer the boundary terms, one could try
to use different species on the boundaries with interactions
satisfying $\tilde{P}^{3/2}_{j,j'}=
\frac{1}{3} (3 {\bf 1}_9+2\tilde{{\bf s}}_{j}\cdot {\bf S}_{j'}-S^{z 2}_{j}),$
where $\tilde{s}^{\mu}$ acts on in the basis 
$\{\ket{S^z=1},\ket{S^z=-1}\}$ and zero elsewhere.

An alternative is trapped spin-$1$ polar molecules
with microwave-induced dipole-dipole interactions instead of tunneling along 
$\hat{z}$. In Ref.~\cite{Brennen:07}, it was shown 
that the AKLT spin lattice model can be realized with $J\approx 13.7$ kHz 
($\Delta E\approx 4.8$ kHz). Boundaries can be loaded with 
spin-$1/2$ species molecules using an additional optical lattice, and the 
necessary interactions could be designed using spectroscopic 
resolvability for the microwave coupling fields between the different species.  
Interactions can be turned off by expanding the lattice along $\hat{z}$ while 
tuning the microwave fields to keep the same interactions but with reduced 
strength.  Then a rightmost molecule can be moved away from 
the bulk using optical tweezers and measured using state-dependent 
photoionization.  The {\tt CPHASE} gates can be generated between a pair of 
molecules in adjacent wires by outcoupling them using optical tweezers and 
using a microwave field to map $\ket{S^z=1}\rightarrow \ket{S^z=0}$.  
Then $H^{\rm int}$ is available using a single microwave field to induce 
the interaction $({\bf 1}_9-S^{z 2})^{\otimes 2}$ \cite{Brennen:07}, 
and measuring the spins in this new basis. 
In both implementations, the initial ground state can be prepared efficiently 
by adiabatically increasing the interaction of $H$ from 
a configuration with antiparallel spins induced by a staggered magnetic field 
as in Ref.~\cite{Cirac:04}.

 {\bf Conclusion.---}
We have described a scheme to perform MQC entirely within the gapped ground 
state of quantum many-body system with a two-body interaction.  
For ground state protection, it is vital that the physical spin states 
are {\it degenerate} energy levels \cite{Brown:07}, meaning that local 
amplitude and phase errors are equally likely.
The constant energy gap $\Delta E$ protects against noise whose 
spectral weight is smaller than it, providing a mechanism for 
anti-aging of the computational resource.

We acknowledge discussions with P. Zoller, A. Micheli, H.J. Briegel, 
M. Lukin, D. Gross, J. Eisert, and M.A. Martin-Delgado.
The work is supported by FWF, and the EU projects (OLAQUI, SCALA, QICS).


\begin{thebibliography}{99}

\bibitem{Jordan:06} 
S.P. Jordan, E. Farhi, and P.W. Shor, Phys. Rev. A {\bf 74}, 052322 (2006).

\bibitem{Raussendorf:01}
R. Raussendorf and H.J. Briegel, Phys. Rev. Lett. {\bf 86}, 5188 (2001). 

\bibitem{FTMQC}
M.A. Nielsen and C.M. Dawson, Phys. Rev. A {\bf 71}, 042323 (2005); 
P. Aliferis and D.W. Leung, Phys. Rev. A {\bf 73}, 032308 (2006); 
R. Raussendorf, J. Harrington, and K. Goyal, New J. Phys. {\bf 9}, 199 (2007).

\bibitem{Bartlett:06}  
S.D. Bartlett and T. Rudolph, Phys. Rev. A {\bf 74}, 040302(R) (2006).

\bibitem{Gross:07a}
D. Gross and J. Eisert, Phys. Rev. Lett. {\bf 98}, 220503 (2007).

\bibitem{Gross:07b}
D. Gross, J. Eisert, N. Schuch, and D. Perez-Garcia, 
Phys. Rev. A {\bf 76}, 052315 (2007).  

\bibitem{AKLT:88}  
I. Affleck, T. Kennedy, E.H. Lieb, and H. Tasaki, Comm. Math. Phys. {\bf 115}, 
477 (1988).

\bibitem{Werner:91}
D.P. Arovas, A. Auerbach, and F.D.M. Haldane, Phys. Rev. Lett. {\bf 60}, 531 
(1988);
F.M. Fannes, B. Nachtergaele, and R.F. Werner, J. Phys. A {\bf 24}, L185 (1991).

\bibitem{Verstraete:04a}
F. Verstraete, M.A. Martin-Delgado, and J.I. Cirac, Phys. Rev. Lett. {\bf 92}, 
087201 (2004).

\bibitem{Fan:04}
H. Fan, V. Korepin, and V. Roychowdhury, Phys. Rev. Lett. {\bf 93}, 227203 
(2004).

\bibitem{Verstraete:04b}
F. Verstraete and J.I. Cirac, Phys. Rev. A {\bf 70}, 060302(R) (2004).

\bibitem{Schon:07}
C. Sch\"on {\it et al.}, Phys. Rev. A {\bf 75}, 032311 (2007).

\bibitem{Yip:03}
S.K. Yip, Phys. Rev. Lett. {\bf 90}, 250402 (2003).

\bibitem{Cirac:04}  
J.J. Garc\'{\i}a-Ripoll, M.A. Martin-Delgado, and J.I. Cirac, 
Phys. Rev. Lett. {\bf 93} , 250405 (2004).

\bibitem{Porto:03}
J.V. Porto {\it et al.}, Phil. Trans. R. Soc. Lond. A {\bf 361}, 1417 (2003).

\bibitem{Zhang:06} 
C. Zhang, S. L. Rolston, and S. Das Sarma, Phys. Rev. A {\bf 74}, 042316 (2006).

\bibitem{Strauch:07}  
F. Strauch {\it et al.}, Phys. Rev. A {\bf 77}, 050304(R) (2008).

\bibitem{Brennen:07}
G.K. Brennen, A.Micheli, and P. Zoller, New J. Phys. {\bf 9}, 138 (2007).

\bibitem{Brown:07} 
K.R. Brown, Phys. Rev. A {\bf 76}, 022327 (2007).


\end{thebibliography}
\end{document}